# Verification and Validation of a Rapid Design Tool for the Analysis of the Composite Y-Joint of the D8 Double-Bubble Aircraft

Evgenia Plaka[a], Stephen P. Jones[b], Brett A. Bednarcyk[c], Evan J. Pineda[c], Richard Li[d], Marianna Maiaru[a]

**Polymer composite joints are critical aerospace components for reinforcing lightweight structures and achieving high eco-efficiency transportation standards. Optimizing complex structural joints is an iterative process. Fast and reliable numerical approaches are needed to overcome the runtime limitations of high-fidelity Finite Element (FE) modeling. This work proposes a computationally efficient approach based on the design tool, HyperX. Verification against FE models and experimental validation are presented for the composite Y-joint in the D8 double bubble fuselage. Results show that the failure load of the Y-joint is predicted within 10% of the experimental failure load recorded. Two parametric studies are performed to study the effects of the curvature of the joint (110° - 160°) and the skin thickness (16ply, 24ply, 32ply) in the failure load predictions using a stress-based failure criterion. The maximum failure load occurred for a Y-joint with 130° curvature. The 32ply skin Y-joint was predicted to have the highest failure load. Results prove the applicability of rapid joint optimization analysis for faster, computationally efficient design.**

## 1. Introduction

Adhesively bonded structural joints are widely used in the aerospace industry [1], [2] due to their high strength-to-weight ratio, amongst other advantages [3]. When joining composite components, adhesive bonding is preferred mainly for two reasons. Firstly, adhesives distribute the load more evenly instead of creating stress concentrations, such as around fasteners, while avoiding continuous fiber breakage, and secondly, fasteners increase overall part count, leading to higher costs [4] - [6]. While unidirectional fiber-reinforced polymer composites (FRPCs) have been used for a range of applications offering high specific strength and stiffness, there has been an increase in the production and use of 3D woven composite materials due to their enhanced out-of-plane (through-thickness) properties amongst other advantages [7] - [10], which are favorable in applications with significant out-of-plane loading. Joints that

[a] Department of Mechanical Engineering, University of Massachusetts Lowell, Lowell, MA 01854

[b] Collier Aerospace, Newport News, Virginia 23606

[c] NASA Glenn Research Center, Cleveland, Ohio 44135

Corresponding author: Marianna Maiaru, Associate Professor, Department of Mechanical Engineering, University of Massachusetts Lowell, 1 University Avenue, Dandeneau Hall 248, Lowell, MA 01854, E-mail: marianna_maiaru@uml.edu

connect a perpendicular structural member to fuselage panels, such as Pi-joints, are examples of applications where out-of-plane loadings define the design. Pi-joints have been extensively studied and characterized in literature; testing indicates that such joints exhibit delamination on the preform legs under tensile loading. Bolts are often used to reinforce the preform legs and prevent delamination [11] - [14].

This work analyzes a novel composite structural joint that connects two fuselage lobes, as shown in Fig. 1. Such a new design, derived from introducing curvature in the preform legs, has been proposed by Aurora Flight Sciences, A Boeing Company (Aurora). The Y-joint is a critical component of the D8 Double Bubble aircraft, an ultra-efficient plane proposed as a result of the NASA N+3 study [15] - [19]. Aurora's current Y-joint joint design comprises a 3D woven composite preform material and unidirectional FRPC skin. According to Aurora's experimental observations [15], [16], the current Y-joint design consistently failed in the joint center region close to the preform-to-skin interface, as shown in Fig. 1.

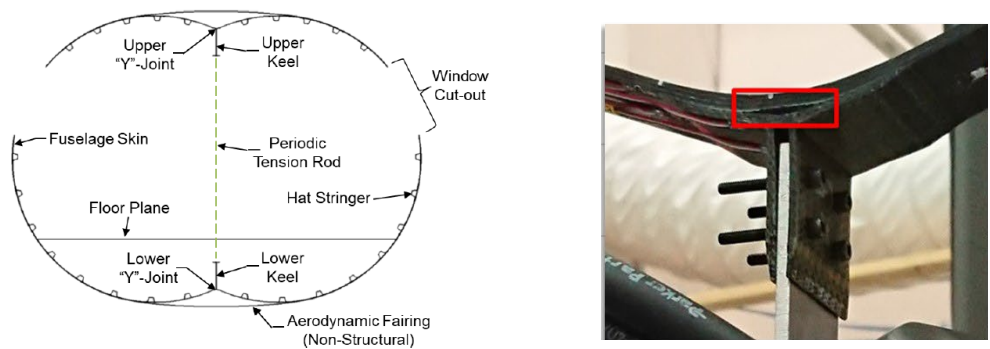

**Fig. 1.** Cross-sectional view of the D8 fuselage (left) and initial failure region in the Y-joint as seen in experiments (right) [16].

Optimizing the current Y-joint configuration requires extensive and iterative structural analyses that require long computational times if high-fidelity models are employed. Design tools based on rapid, lower-fidelity models can be used for optimization upon their verification against high-fidelity validated models. HyperX is a new commercial software for joint design based on enhanced analytical solutions, which includes numerous joint configurations and extended capabilities from its precursor software, HyperSizer [20]. The use of HyperSizer has been extensively documented in the literature. Zhang et al. proved HyperSizer capability to compute 3D stress states in adhesively bonded composite joints [21] for single-lap, double-lap, scarfed, and stepped joints. Stapleton et al. evaluated the tool's local stress field predictions for double lap joints and found good agreements against a high-fidelity finite element model [22]. Additionally, failure prediction capabilities have been developed in HyperX for all the joint configurations supported by the software. Yarrington et al. evaluated HyperX accuracy and consistency in predicting initial failure for the bonded doubler, stepped bonded doubler, and single-lap joint configurations [23]. Yarrington et al. also developed the virtual crack closure technique for the growth prediction of a preexisting crack [24], and Jones et al. evaluated its applicability on HyperX's adhesively bonded joint configurations [25]. Currently, no Pi-joint or Y-joint configurations are implemented in HyperSizer or HyperX. However, Plaka et al. [26], [27] have proposed a methodology for using HyperX to analyze Pi- and Y-joints.

This work establishes a rapid and comprehensive method to optimize Aurora's D8 Y-joint using HyperX. First, HyperX [28] is verified against traditional finite elements in Abaqus. Then, an approach to decompose the Y-joint into two simpler joints for analysis in HyperX is proposed by applying boundary conditions that replicate the effect of the curvature in the preform legs. Results, analyzed in the failure region of the Y-joint (see Fig. 1), represented in HyperX as a doubler joint, show good agreement between HyperX and finite elements. The bonded doubler configuration is shown in this paper as part of the verification of HyperX because the region of interest (region where the Y-joint failed, see Fig. 1) is part of the doubler configuration. Results show the agreement between finite element and HyperX results, demonstrating HyperX as a viable solution for rapid analysis of the Y-joint configuration for this study. Lastly, HyperX is validated against Aurora's experimental results. One of the configurations Aurora tested is reconstructed in HyperX, and the failure loads predicted for that configuration using stress-based failure criteria were compared to the experimental failure pull-off load measured. Lastly, two parametric studies were conducted in HyperX. The impact of two parameters, the joint curvature and skin thickness (one of the joint adherends), on the predicted failure load are investigated. The joint curvatures are varied between 110° and 160°, while the three different laminated skins explored are 16-ply, 24-ply, and 32-ply laminated skin. Failure loads for both parametric studies were obtained using a stress-based criterion that considers peel, longitudinal, transverse shear, and axial stresses to approximate delamination-related failure in the adherend.

The analysis proposed in this paper is part of an Integrated Computational Materials Engineering (ICME) approach to design and optimize the composite Y-joints and composite acreage panels used in the Aurora D8 aircraft [15] - [17], [29], [30]. The ICME framework links material models, structural models, and experiments at multiple length scales [31] - [34] and aims to provide a better understanding of the tie between product design and composite manufacturing through experimentally validated simulations [35]. Specifically, the manufacturing-microstructure correlation and its effect across the higher length scales are established through atomistically-informed process modeling simulations

[36], [37] embedded into Finite Element (FE) micromechanics models [38] - [41]. Virtually cured micromechanics models are linked to meso- and macro-scale models to determine strength allowables affected by curing-induced residual stress [42]. Results from each scale will be used to modify the cure cycle, toughen specific regions of the model, and selectively modify the composite layup leveraging the proposed rapid optimization procedure highlighted in this manuscript.

This paper is organized as follows: Section 2 shows HyperX verification against high-fidelity 3D FE simulations, and Section 3 details the validation of the proposed approach against experimental results from Aurora. Section 4 proposes a sensitivity analysis of the failure analysis based on analyzing variations in the Y-joint curvature and thickness of the adherends. Lastly, Section 5 summarizes conclusions from the verification, validation, and sensitivity analyses.

## 2. Verification of HyperX with Abaqus for The Y-joint

HyperX is a structural analysis and sizing software that includes stress analysis and optimization of joints. The stress analysis is based on Mortensen's unified approach [43], [44], and it is developed further to consider the through-thickness interlaminar peel and shear stresses for laminated adherends. The joint configurations supported by HyperX are shown in Fig. 2. Since HyperX does not currently include a Y-joint configuration directly, a new methodology has been proposed and used to analyze such joints.

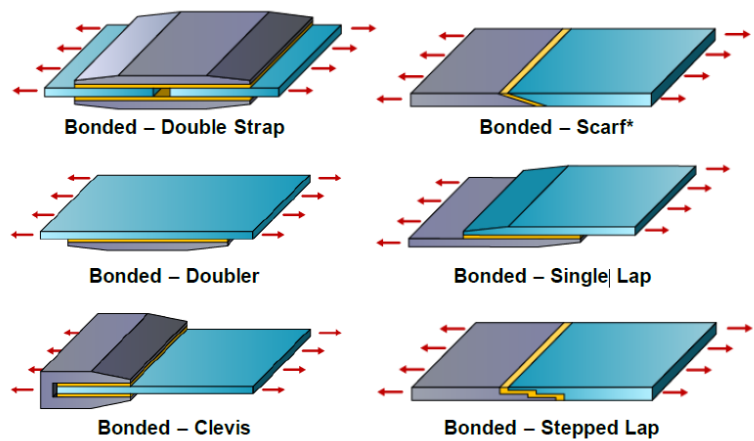

**Fig. 2** Joint configurations supported by HyperX [25].

Plaka et al. [26], [27] proposed that a combination of two joint configurations available in HyperX can be used as an equivalent to the Y-joints for the purposes of this work. Fig. 3 shows the regions in the Y-joint that were analyzed in HyperX as two different joint configurations. Region 1 was modeled as a bonded clevis, while Region 2 was modeled as a bonded doubler. As previously mentioned, according to experimental tests from Aurora, the Y-joint consistently failed in the joint center region close to the preform-to-skin interface, under the preform uprights [15], [16]. Thus, this study places emphasis on this region, where the failure was observed. Since the region of interest is part of the bonded doubler joint configuration, only results for the doubler are shown in this section to summarize the verification between HyperX and Abaqus. A detailed methodology and verification analysis for a Pi-joint, as well as a Y-joint has been established [26], [27].

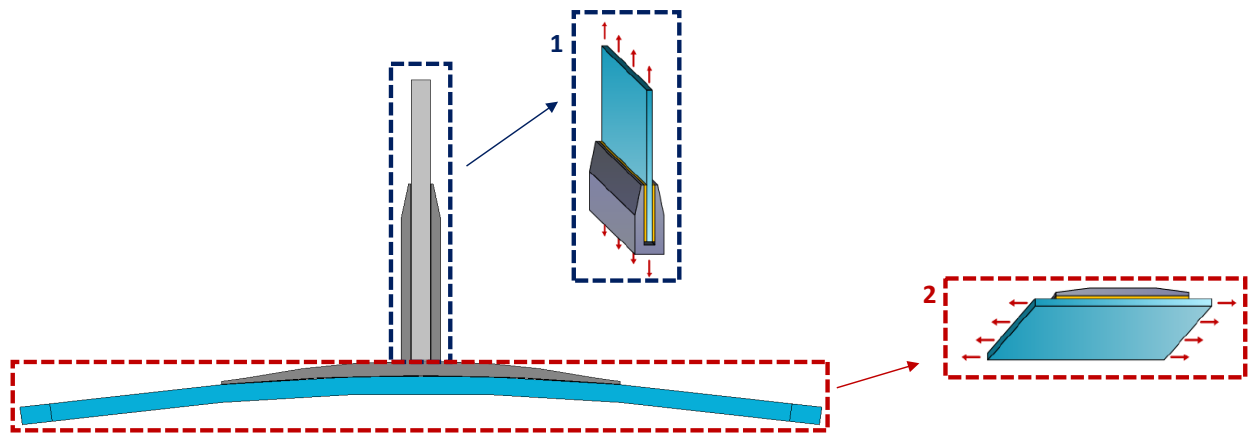

**Fig. 3** Regions in the Abaqus model that were analyzed as two separate joints in HyperX (region 1: bonded clevis, region 2: bonded doubler).

The Y-joint as modeled in Abaqus is shown in Fig. 4. The angle of curvature was 175.2°. The thickness of the skin was 5.8mm and consisted of 32 plies in the quasi-isotropic laminate with stacking sequence ($[0/\pm45/90]_{4S}$). The woven preform was modeled as a homogeneous solid and given effective properties, and the preform base had a thickness of 3.81 mm. The adhesive layers had a thickness of 0.1016 mm and were modeled as homogenous solid elements. The 2D Y-joint was modeled using plane strain CPE4 and CPE3 solid elements in a linear elastic analysis. The composite material properties for the laminated skin and woven preform were based on AS4 fiber and RTM6 matrix constituents, provided by NASA [45]. Since this section sought to compare the two software rather than determine their absolute accuracy, rule of mixture was used to determine all the composite properties. The aluminum bar was made of Al 7075-T6, while the adhesive used was FM300K epoxy film adhesive 293 gsm. All material properties are given in Table 1 below. The equivalent Y-joint geometry of region 2 (see Fig. 3) modeled in HyperX as the bonded doubler joint configuration is shown in Fig. 5.

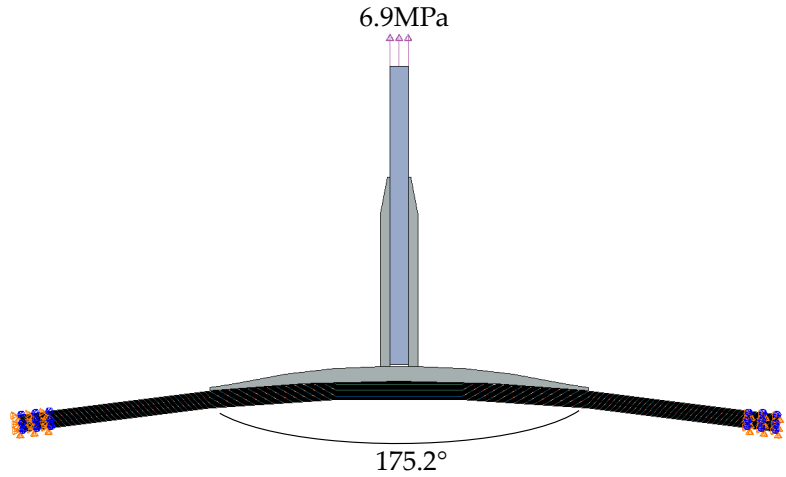

**Fig. 4.** Y-joint model with 175.2° angle of curvature as modeled in Abaqus.

**Table 1** Elastic properties for the AS4/RTM6 3D woven preform, AS4/RTM6 0° plies, Al 7075-T6 aluminum bar and FM 300K film adhesive.

|  | AS4/RTM6 3D Woven Preform | AS4/RTM6 0° ply | Al 7075-T6 | FM 300K Epoxy Film Adhesive 293 gsm |
|---|---|---|---|---|
| $E_1$ (GPa) | 59.32 | 139.8 | 71.71 | 2.357 |
| $E_2$ (GPa) | 59.12 | 5.610 | 71.71 | 2.357 |
| $E_3$ (GPa) | 9.500 | 5.610 | 71.71 | 2.357 |
| $\nu_{12}$ | 0.037 | 0.266 | 0.300 | 0.300 |
| $\nu_{13}$ | 0.318 | 0.266 | 0.300 | 0.300 |
| $\nu_{23}$ | 0.305 | 0.318 | 0.300 | 0.300 |
| $G_{12}$ (GPa) | 2.061 | 2.430 | 27.58 | 0.907 |
| $G_{13}$ (GPa) | 2.047 | 2.430 | 27.58 | 0.907 |
| $G_{23}$ (GPa) | 4.074 | 2.120 | 27.58 | 0.907 |

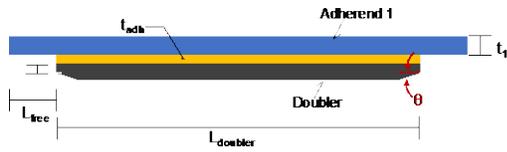

| Geometric Parameter | Abbreviation | Value |
|---|---|---|
| Adherend 1 Thickness (mm) | $t_1$ | 4.57 |
| Doubler Thickness (mm) | $t_{doubler}$ | 3.81 |
| Adhesive Thickness (mm) | $t_{adh}$ | 0.10 |
| Doubler Length (mm) | $L_{doubler}$ | 101.6 |
| Taper Angle (°) | $\theta$ | 26.56 |
| Final Thickness (mm) | $t_{final}$ | 1.27 |
| Free Length (mm) | $L_{free}$ | 43.18 |

**Fig. 5.** Geometric parameters for bonded doubler in HyperX for the Y-joint with 175.2° angle of curvature.

The boundary conditions applied to the two individual HyperX joint concepts (clevis and doubler) are critical and must be defined appropriately to capture the correct mechanics of an effective Y-joint. In Abaqus, the Y-joint model was held fixed in both the legs of the skin, and the aluminum bar was loaded in the y-direction by a pressure of -6.9 MPa. Using the method proposed and described in detail by Plaka et al. [26], [27], the boundary conditions (locations shown in Fig. 6) assigned on the bonded doubler in HyperX are shown in Table 2.

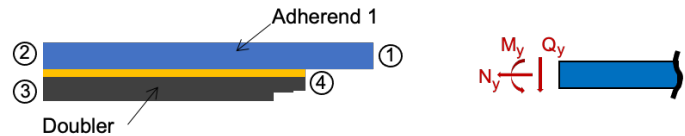

**Fig. 6.** Boundary condition locations for the bonded doubler (left) and sign convention (right) in HyperX.

**Table 2** Boundary conditions for the Y-joint doubler.

| Location | Axial | Transverse | Moment | Shear |
|---|---|---|---|---|
| 1 | Fixed | Fixed | Fixed | Fixed |
| 2 | Fixed | Free | Free | Free |
| 3 | $N_y = 0.704$ N/mm | $Q_y = 16.7$ N/mm | $M_y = -5.97$ N · mm/mm | Free |
| 4 | Free | Free | Free | Free |

The peel and shear stresses are considered critical parameters for adhesives, therefore they have been used in this work to compare Abaqus and HyperX. Stresses between the preform base and the skin are plotted in half of the adhesive of the doubler region because of symmetry. All stresses shown in this section are plotted against the distance along the adhesive path (a schematic of the doubler region shows the path in the plots). Fig. 7 shows the peel and shear stresses in the adhesive of the bonded doubler in the Y-joint. As previously stated, failure was observed right under the preform uprights region, which is included in the plots in Fig. 7. It is shown that HyperX can capture the first-order physics of the Y-joint behavior under tensile loading. The difference in stresses in the region where the taper is modeled (closer to x = 40 mm) can be attributed to the effects of the adjacent face sheet ply, which HyperX cannot consider because the adhesive stresses are calculated based on homogenized adherend laminate properties. Those differences are acceptable for the purpose of this paper because that region is not part of the area of failure. Additionally, in HyperX adhesives are modeled as springs and the adherends as shell elements. However, the good qualitative agreement indicates that HyperX can readily be used for design through appropriate correlation with experimental joint strength data.

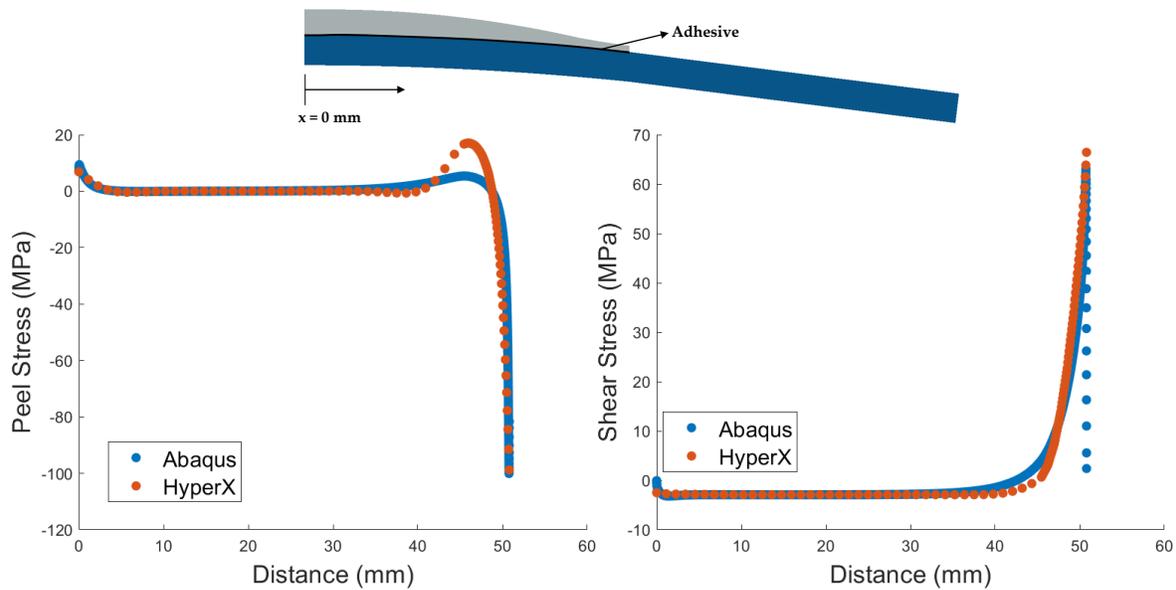

**Fig. 7.** Peel stress (left) and shear stress (right) along the adhesive of the Y-joint doubler as predicted by Abaqus and HyperX.

Fig. 8 confirms that the HyperX trend predictions of the peak shear and peel stresses in the adhesive of the doubler for the region of interest match the finite element prediction trends. It is evident that the curvature in the joint results in more differences in the stress results between the two software. Nevertheless, since the optimization process requires various iterations of different designs and multiple parametric studies, the rapid design tool capability will be mainly used for its trend predictive capabilities.

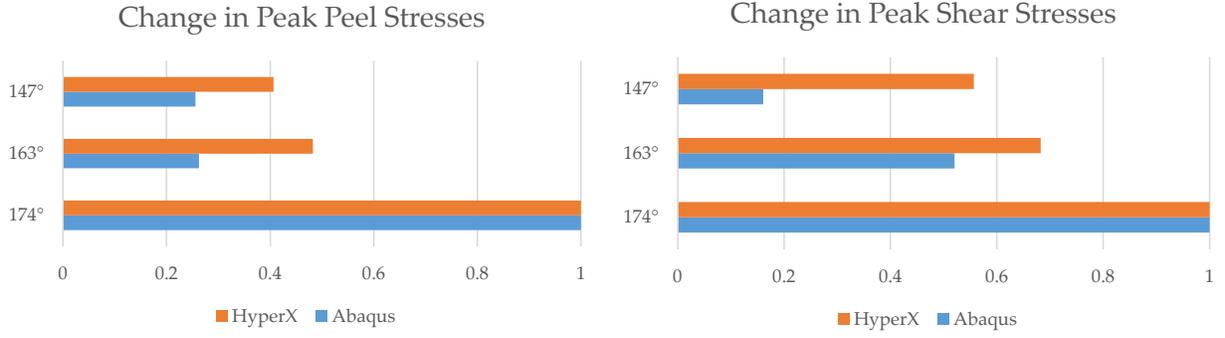

**Fig. 8.** Peak peel stress (left) and peak shear stress (right) along the adhesive of the Y-joint doubler, in the region of interest, as predicted by Abaqus and HyperX for three different curvatures.

## 3. Validation of HyperX

Using the method verified in the previous section, the Y-joint model was also validated against experimental results. Failure prediction capabilities have been developed in HyperSizer and included in HyperX for the joint configurations currently supported [23]- [25]. An initial investigation into the failure load prediction capability of HyperX for the Y-joint was done by Plaka et al. [27] and it was concluded that, of the stress-based failure criteria available in HyperX, interactive failure criteria are best for the present application since previous work on the Y-joint in HyperX has shown that the failure is driven both by shear and peel stresses. In previous work two interactive stress-based failure criteria were determined to be the most appropriate to use in this analysis [27]. The two failure criteria used, proposed by Tong [46] - [48], are defined by the two equations given below [49]:

Delamination, Tong, Peel, Transverse Shear & Axial 3

$$\left(\frac{\sigma_1}{F_1^{tu}}\right)^2 + \left(\frac{\sigma_3}{F_3^{tu}}\right)^2 + \left(\frac{\tau_{13}}{F_{13}^{su}}\right)^2 = 1 \qquad (1)$$

Delamination, Tong, Peel, Transverse Shear & Axial 4

$$\left(\frac{\sigma_1}{F_1^{tu}}\right)^2 + \left(\frac{\sigma_3}{F_3^{tu}}\right) + \left(\frac{\tau_{13}}{F_{13}^{su}}\right)^2 = 1 \qquad (2)$$

where $F_1^{tu}$, $F_3^{tu}$, and $F_{13}^{su}$ are the ultimate normal stress allowable in the 1 direction, the ultimate normal stress allowable in the 3 direction, and the out-of-plane shear stress allowable in 13 ply coordinates respectively; and $\sigma_1$, $\sigma_3$, and $\tau_{13}$ are the normal stress in 1 direction, normal stress in 3 direction, and shear stress in 13 ply coordinates respectively.

The HyperX model was constructed to match configuration B from Aurora's experiments [16]. Using images of the joint from [16] a simple pixel analysis method was performed to estimate all dimensions of the Y-joint. Using that geometry and two layers of film adhesive, HyperX was used to predict the failure pull-off load of the Y-joint. The geometry of the bonded doubler used for the validation is shown in Fig. 9. The material used for the composite preform was T800/3900 3D weave, while the skin was made of T800/3900 unidirectional (UD) plies in a [0/45/-45/90]$_{3S}$ stacking sequence. The adhesive used was FM300M 146 gsm film adhesive. Adhesive nonlinear analysis was conducted [49]. The skin and preform material properties used for this analysis are shown in Table 3. The material properties used for the adhesive material are given in Table 4.

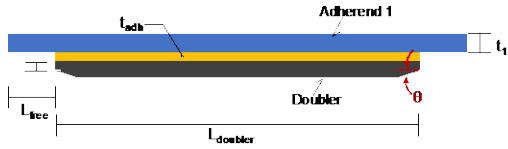

| Geometric Parameter | Abbreviation | Value |
|---|---|---|
| Adherend 1 Thickness (mm) | $t_1$ | 4.57 |
| Doubler Thickness (mm) | $t_{doubler}$ | 3.81 |
| Adhesive Thickness (mm) | $t_{adh}$ | 0.25 |
| Doubler Length (mm) | $L_{doubler}$ | 86.0 |
| Taper Angle (°) | $\theta$ | 15.9 |
| Final Thickness (mm) | $t_{final}$ | 0.25 |
| Free Length (mm) | $L_{free}$ | 142.2 |

**Fig. 9.** Geometry of bonded doubler in HyperX equivalent of Y-joint in configuration B [16].

Using two different stress-based interactive delamination criteria for the adherends, two different failure loads were predicted, which gave an upper and a lower failure load limit. Local stresses are taken from the region of the doubler between x = 0 mm and x = 20.3 mm (see Fig. 10), to avoid the evaluation of oscillating stresses at the end of the doubler region (discussed in the verification section) in the calculation of margins of safety. HyperX evaluates failure criteria at all points in the considered regions of the joint, and the failure load has been evaluated as the lowest load that first causes the failure criterion to be exceeded at some point in the joint.

Table 3 Elastic material properties and stress allowable for T800/3900 plies and 3D weave.

|  | T800/3900 Ply (0°) | T800/3900 3D Weave |
|---|---|---|
| $E_1^t$ (GPa) | 148.2 | 68.95 |
| $E_2^t$ (GPa) | 8.549 | 67.98 |
| $\nu_{12}^t$ | 0.332 | 0.032 |
| $E_1^c$ (GPa) | 130.3 | 63.64 |
| $E_2^c$ (GPa) | 8.481 | 61.50 |
| $\nu_{12}^c$ | 0.339 | 0.044 |
| $G_{12}$ (GPa) | 3.937 | 3.592 |
| $G_{13}$ (GPa) | 3.937 | 2.110 |
| $G_{23}$ (GPa) | 2.910 | 2.110 |
| $F_1^{tu}$ (MPa) | 3006 | 1034 |
| $F_2^{tu}$ (MPa) | 61.57 | 923.9 |
| $F_1^{cu}$ (MPa) | 1779 | 668.1 |
| $F_2^{cu}$ (MPa) | 215.8 | 603.3 |
| $F_{12}^{su}$ (MPa) | 69.64 | 79.29 |
| $F_{13}^{su}$ (MPa) | 69.64 | 74.46 |
| $F_{23}^{su}$ (MPa) | 38.69 | 70.05 |
| $F_3^{tu}$ (MPa) | 60.12 | 89.77 |

Table 4 Elastic material properties and stress allowables for FM300M 146 gsm film adhesive.

|  | $E^t$ (GPa) | $E^c$ (GPa) | $G$ (GPa) | $\nu^t$ | $\nu^c$ | $n$ | $F0.2$ (kPa) | $F_{peel}^{adh}$ (MPa) | $F_{shear}^{adh}$ (MPa) |
|---|---|---|---|---|---|---|---|---|---|
| FM300M 1.44 Pa | 2.062 | 2.062 | 0.793 | 0.300 | 0.300 | 17.10 | 30.08 | 78.82 | 39.41 |

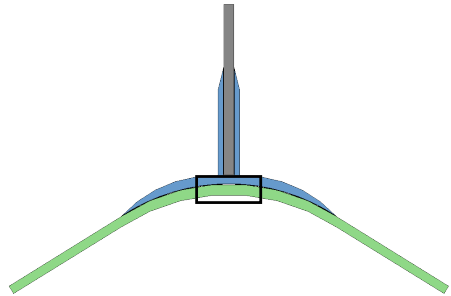

**Fig. 10.** Region of Y-joint that was considered for failure analysis (black box).

The two failure loads predicted by HyperX were 19.25 kN and 17.28 kN according to Tong, Peel, Transverse Shear & Axial 3 and 4 respectively. Both of those values lie within 10% of Aurora's average experimental results (the experimental results showed an average ultimate failure load of 18.68 kN [16]). This indicates that HyperX, as applied to the curved Y-joint, is capable of accurately predicting failure loads and optimizing the Y-joint.

## 4. Parametric Studies

Parametric studies can reveal crucial sensitivities that can guide the optimization process. In this study, two major parametric studies were completed. The parameters explored were the curvature of the joint and the thickness of laminated skin. Upon verifying that HyperX can predict the trends in stress peaks [26], failure load predictions were made using HyperX. Using the same analysis process as described in the validation (see section 3), the material system used was T800/3900 for the adherends and FM300M 1.44 Pa for the adhesive. Non-linear adhesive analysis was once again considered and the region in which the local stresses are taken to calculate the failure remain the same (see Fig. 10). While failure loads were predicted according to Tong 3 (Eq. 1) and Tong 4 (Eq. 2), in this paper the most conservative failure criterion of the two was chosen to present the predicted failure load results.

### 4.1 Curvature Failure Load Predictions

A range of different curvatures was studied starting from 110° and reaching 160°. The HyperX boundary conditions are set accordingly using the method discussed in the verification process, given in section 2. The material properties used are given in Table 3 and Table 4. The laminated skin had a thickness of 5.08 mm, and the layup was $[0/45/-45/90]_{4s}$. Two layers of FM300M adhesive were used, with a total thickness of 0.254 mm. The preform thickness was set to 3.81 mm.

Results obtained from HyperX for the sensitivity study of the effect of curvature in the failure load on the Y-joint are shown in Fig. 11. Predictions show that at 130° angle of curvature of the Y-joint, a maximum failure pull-off load is observed. At curvatures below and above 130°, in the range of 110°-160°, the predicted failure load decreases.

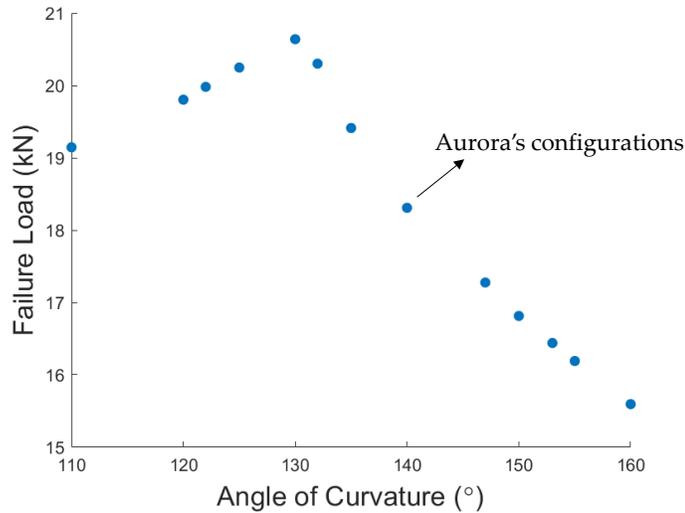

**Fig. 11.** Failure load predictions in HyperX for Y-joints with different curvatures using the Tong 4 failure criterion, (2).

### 4.2 Skin Thickness Failure Load Predictions

The effects of a thin and a thick laminated skin were investigated using a 16 ply skin, 24 ply skin, and 32 ply skin. The laminate layup was $[0/45/-45/90]_{nS}$ with a ply thickness of 0.1778 mm. The material properties remain the same (see Table 3 and Table 4). The curvature of the Y-joints for this study is held constant at 146.6°. Similar to the curvature sensitivity study, the total thickness of the adhesive was 0.254 mm, and the thickness of the preform was 3.81 mm.

Fig. 12 shows the predicted failure loads for three Y-joints of curvatures 146.6°, and 16, 24 and 32 plies in the skin. HyperX predicts that the failure load increases with increasing skin thickness. The 24ply skin Y-joint model is the same model used for the validation of HyperX, matching configuration B from Aurora's experiments [16]. The 16ply skin model corresponds to configuration E from Aurora's tests, and the 32ply skin model to the configuration H. Even though configuration E's average failure load matches within 10% with the predicted failure load of the 16ply skin model, Aurora reported that the failure load results measured for configurations E and H are tainted due to manufacturing defects [16].

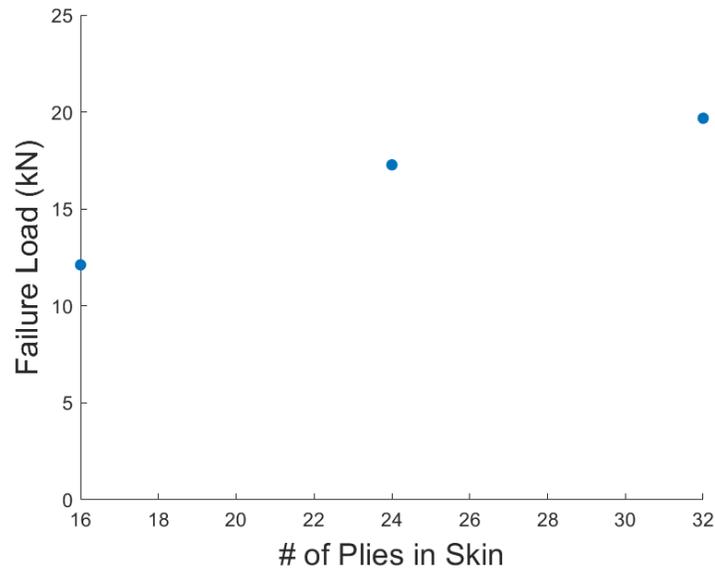

**Fig. 12.** Failure load predictions in HyperX for Y-joints with skin thicknesses of 16 ply, 24 ply and 32 ply using the Tong 4 failure criterion, (Eq. 2).

## 5. Conclusions

This paper presents insights into the optimization process of a composite Y-joint used in a multi-lobe aircraft fuselage and shows the applicability of the rapid joint design tool, HyperX, for analyzing the Y-joint preform uprights where failure is expected. HyperX has been verified against the finite element analysis software Abaqus and good agreement between the two software has been shown. The limits of applicability of the proposed approach have also been highlighted.

HyperX was used to predict the failure load, using an interactive stress-based criterion, of one of the Y-joint configurations tested by Aurora Flight Sciences and was shown to be within 10% of the average measured value.

This work proposed an approach to explore trends in failure load predictions based on sensitivity studies in HyperX. The curvature and skin thickness in the Y-joint showed the most significant impact on the stress distribution [26] and were chosen to investigate the prediction of failure loads in the joint based on the existing stress-based failure criterion. The curvature sensitivity study showed a maximum failure load seen when the Y-joint has a curvature angle of 130°. In the range investigated, the failure load predicted decreased for curvature angles below and above 130°. HyperX predicted realistic trends in failure loads with varying skin thickness, showing an increase in the failure load with increasing skin thickness.

Future work will include an in-depth study of the prediction of failure loads using energy-based failure criteria to capture the failure propagation and evolution, as well as the integration of this macroscale analysis in a multiscale iterative framework, where various design parameters will be tested for the optimization of the Y-joint, including the effect of residual stress on design allowables.

## 6. Acknowledgments

The Composites Technology for Exploration (CTE) project under NASA Space Technology Mission Directorate (STMD), Game Changing Development (GCD) program, supported the NASA authors. The authors would like to acknowledge the support of the NASA Transformational Tools and Technologies (TTT) project within the Aeronautics Research Mission Directorate under NRA cooperative grant 80NSSC21M0104.